\documentclass[]{article}
\usepackage{amsmath}
\usepackage{icrctc07}

\def \mal   {Malarg\"{u}e}
\def \pao   {Pierre Auger Observatory }
\def \opa   {Observatorio Pierre Auger }

\title{Measurement of Aerosols at the \pao}
\shorttitle{Measurement of Aerosols at the \pao}
\authors{
  S.Y.~BenZvi, F.~Arqueros, R.~Cester, M.~Chiosso, B.M.~Connolly, B.~Fick,
  A.~Filip\v{c}i\v{c}, B.~Garc\'{\i}a, A.~Grillo, F.~Guarino, M.~Horvat,
  M.~Iarlori, C.~Macolino, M.~Malek, J.~Matthews, J.A.J.~Matthews, D.~Melo,
  R.~Meyhandan, M.~Micheletti, M.~Monasor, M.~Mostaf\'{a}, R.~Mussa,
  J.~Pallotta, S.~Petrera, M.~Prouza, V.~Rizi, M.~Roberts, J.R.~Rodriguez~Rojo,
  D.~Rodr\'{\i}guez-Fr\'{\i}as, F.~Salamida, M.~Santander, G.~Sequeiros,
  P.~Sommers, A.~Tonachini, L.~Valore, D.~Verberi\v{c}, E.~Visbal,
  S.~Westerhoff, L.~Wiencke, D.~Zavrtanik, M.~Zavrtanik, for the Pierre Auger
  Collaboration$^1$
  }
\shortauthors{Segev BenZvi and et al}
\afiliations{
  $^1$\opa, Av. San Mart\'{\i}n Norte 304 (5613) \mal, Argentina
}
\email{sybenzvi@phys.columbia.edu}

\abstract{

  The air fluorescence detectors (FDs) of the \pao are vital for the
  determination of the air shower energy scale.  To compensate for variations
  in atmospheric conditions that affect the energy measurement, the Observatory
  operates an array of monitoring instruments to record hourly atmospheric
  conditions across the detector site, an area exceeding 3,000 km$^2$.  This
  paper presents results from four instruments used to characterize the aerosol
  component of the atmosphere: the Central Laser Facility (CLF), which provides
  the FDs with calibrated laser shots; the scanning backscatter lidars, which
  operate at three FD sites; the Aerosol Phase Function monitors (APFs), which
  measure the aerosol scattering cross section at two FD locations; and the
  Horizontal Attenuation Monitor (HAM), which measures the wavelength
  dependence of aerosol attenuation.

}

\begin{document}
\maketitle

\section{Introduction}

  The \pao in \mal, Argentina employs four fluorescence detector (FD)
  telescopes to obtain calorimetric estimates of air shower energies.  The
  atmosphere, which acts as the calorimeter, is constantly in flux,
  so aerosol conditions are measured hourly at each FD location and stored in
  an offline database for use in the reconstruction of showers.

  The aerosol parameters of interest are those that affect light attenuation
  and scattering: $\alpha(h)$, the aerosol extinction coefficient, and
  $\tau(h)$, the vertical aerosol optical depth (VAOD), which is the integral
  of $\alpha(h)$ from the ground to altitude $h$; $P(\theta)$, the normalized
  differential scattering cross section, or phase function, as a function of
  scattering angle $\theta$; and the wavelength dependence of aerosol
  scattering.  We describe measurements of $\tau(h)$ and its wavelength
  dependence by the CLF, lidars, and HAM, and observations of $P(\theta)$ by
  the APFs.

\section{Optical Depth Measurements}

  \subsection{Central Laser Facility}

    The CLF produces calibrated laser ``test beams'' from its location in the
    center of the Auger surface detector~\cite{Fick:2006yn,Wiencke:2006hg}.
    Located between $26$ and $39$~km from the FDs, the CLF uses a pulsed laser
    operating at $355$~nm.  The atmosphere scatters nearly equal amounts of
    laser light toward each FD eye, where it is recorded by the FDs.  With a
    nominal energy of $7$~mJ per pulse, the light produced is roughly equal to
    the amount of scintillation light generated by a $10^{20}$~eV shower.
    
    \begin{figure*}[ht]
      \begin{minipage}[t]{0.49\textwidth}
        \includegraphics*[width=\textwidth,clip]{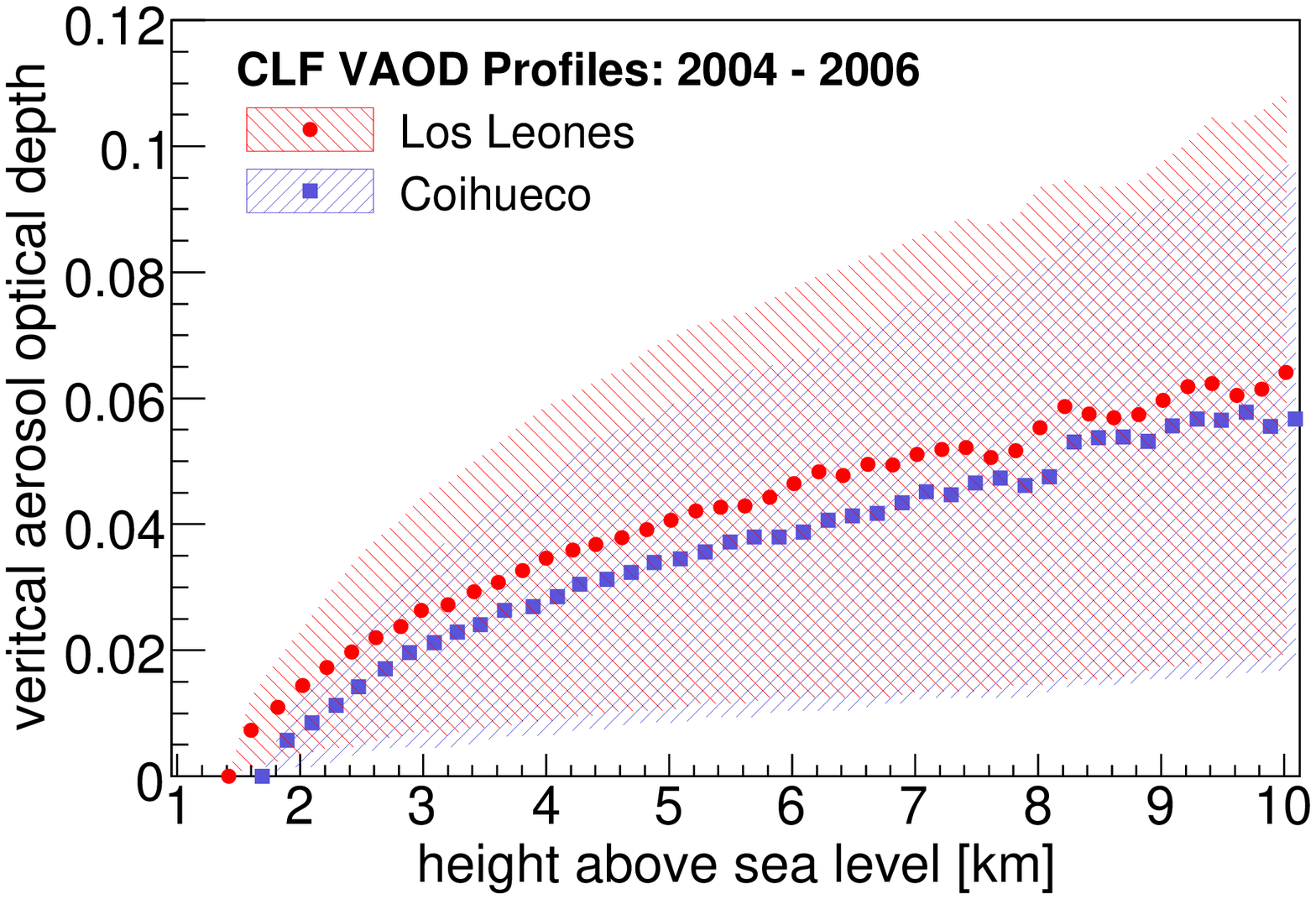}
        \caption{\label{fig:clfVaodErr} CLF VAOD measurements at Los Leones and
                 Coihueco, 2004 - 2006, showing mean VAOD profiles and their
                 68\% confidence limits.}
      \end{minipage}
      \hfill
      \begin{minipage}[t]{0.49\textwidth}
        \includegraphics*[width=\textwidth,clip]{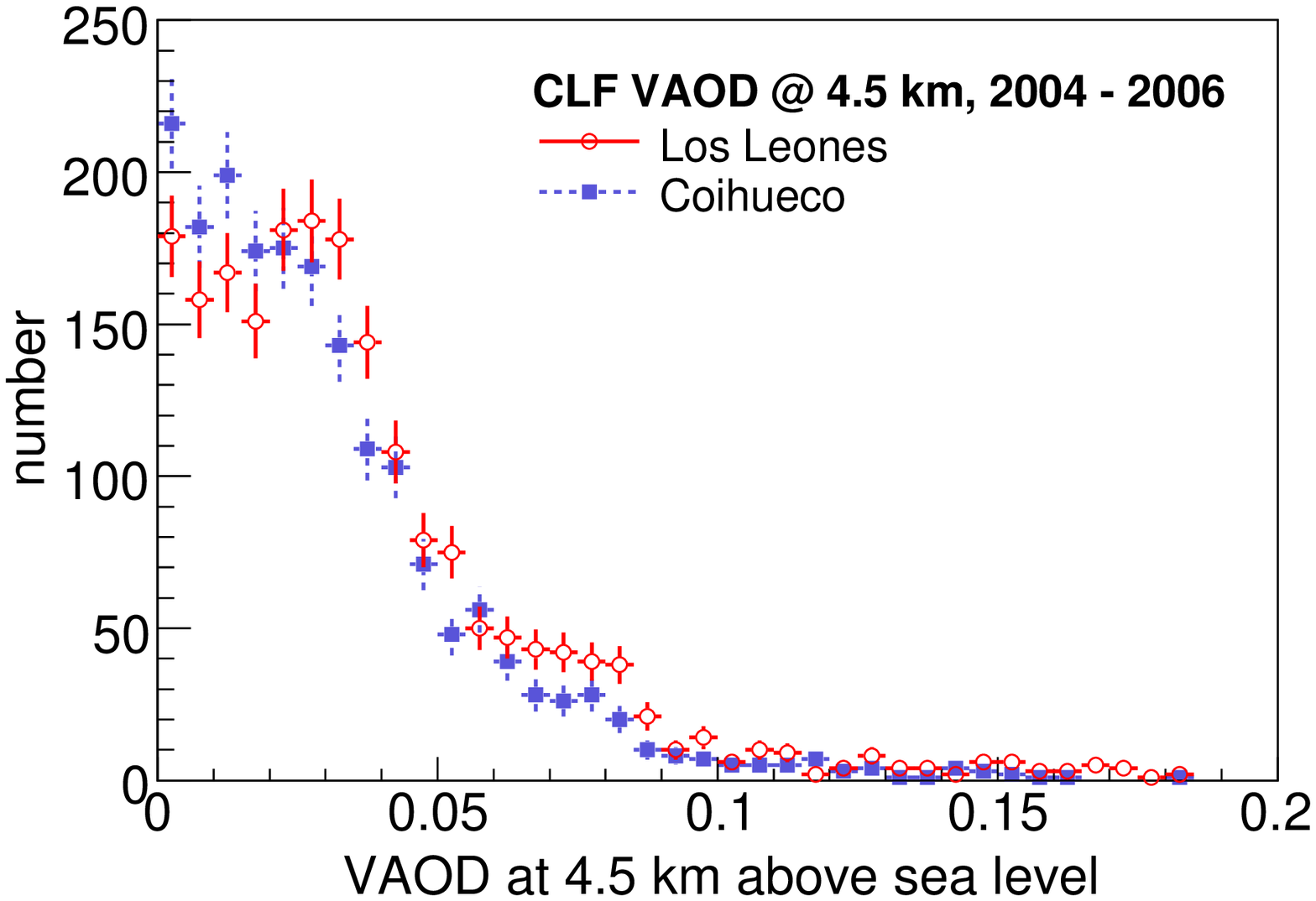}
        \caption{\label{fig:clfVaod45} VAOD at $4.5$~km.  Los Leones, at an 
                 altitude $\sim280$~m below Coihueco, observes a systematically
                 larger VAOD.}
      \end{minipage}
    \end{figure*}

    Among the many measurements provided by the CLF test beams are hourly
    observations of $\alpha(h)$ and $\tau(h)$.  The laser fires sets of 50
    shots every quarter-hour, and the track profiles are averaged to obtain
    hourly aerosol measurements.  The CLF estimates the aerosol content using
    an iterative procedure that does not require absolute photometric
    calibrations of the FD or laser.  The procedure begins by normalizing
    hourly average laser profiles $L(h)$ by an ``aerosol-free'' reference
    profile $L_\mathrm{ref}(h)$ measured during extremely clear conditions.
    The initial VAOD estimate is then

    \begin{equation}\label{eq:clf_init_vaod}
      \tau_i(h) 
        = -\frac{\ln{L(h)} - \ln{L_\mathrm{ref}(h)}}{1 + \csc{\epsilon(h)}}
    \end{equation}

    where $\epsilon$ is the elevation angle to the track point at altitude $h$.
    An estimate $\alpha_i(h)$ comes from the slope of the VAOD, and is used to
    iteratively correct the light profile for aerosol scattering.  At the end
    of the calculation, the final $\alpha(h)$ is normalized by the VAOD at a
    height $h_c$ above the bulk of aerosols:

    \begin{equation}\label{eq:clf_bdry_cond}
      \int_{h_\text{ground}}^{h_c}\alpha(h)dh = \tau_i(h_c)
    \end{equation}

    The final VAOD is found by integrating $\alpha(h)$.
    Figures~\ref{fig:clfVaodErr} and \ref{fig:clfVaod45} depict the VAOD
    distribution recorded at Los Leones and Coihueco between 2004 and 2006.
    Typically, the VAOD at $4.5$~km is 0.03, with statistical uncertainties of
    $\sim0.01$.

    The uncertainties in each VAOD measurement are dominated by statistical
    fluctuations in the hourly average light profiles, although systematic
    effects due to the FD and laser relative calibrations and the choice of
    aerosol-free reference nights are also present.  The last effect has been
    checked with a separate analysis method that uses CLF laser simulations and
    a simple two-parameter exponential model of the aerosol density.
    The results, which are independent of clear-night calibrations,
    closely match the standard CLF estimates (Figure~\ref{fig:clfLaserCorr}).

    The CLF can also detect clouds, which appear as sharp steps in the VAOD
    profile.  When a cloud is present, the lowest base height $h_\mathrm{base}$
    is recorded.  For heights $h\leq h_\mathrm{base}$, $\alpha(h)$ and
    $\tau(h)$ are considered valid for air shower analysis.

  \subsection{Elastic Backscatter Lidars}

    In addition to the CLF, the observatory employs three scanning elastic
    lidar stations, with a fourth under construction, to record $\alpha(h)$ and
    $\tau(h)$ at every FD site~\cite{BenZvi:2006xb}.  Each station has a
    steerable frame mounted with a pulsed $351$~nm laser, three parabolic
    mirrors, and photomultiplier recorders.  The station at Los Leones also
    includes a separate, vertically-staring Raman lidar test system, which can
    detect aerosols and the relative concentration of N$_2$ and O$_2$ in the
    atmosphere.  Since the lidar hardware and measurement techniques are
    independent of the CLF, the two systems have essentially uncorrelated
    systematic uncertainties.

    \begin{figure*}[ht]
      \begin{minipage}[t]{0.49\textwidth}
        \begin{center}
        \includegraphics*[width=0.75\textwidth,clip]{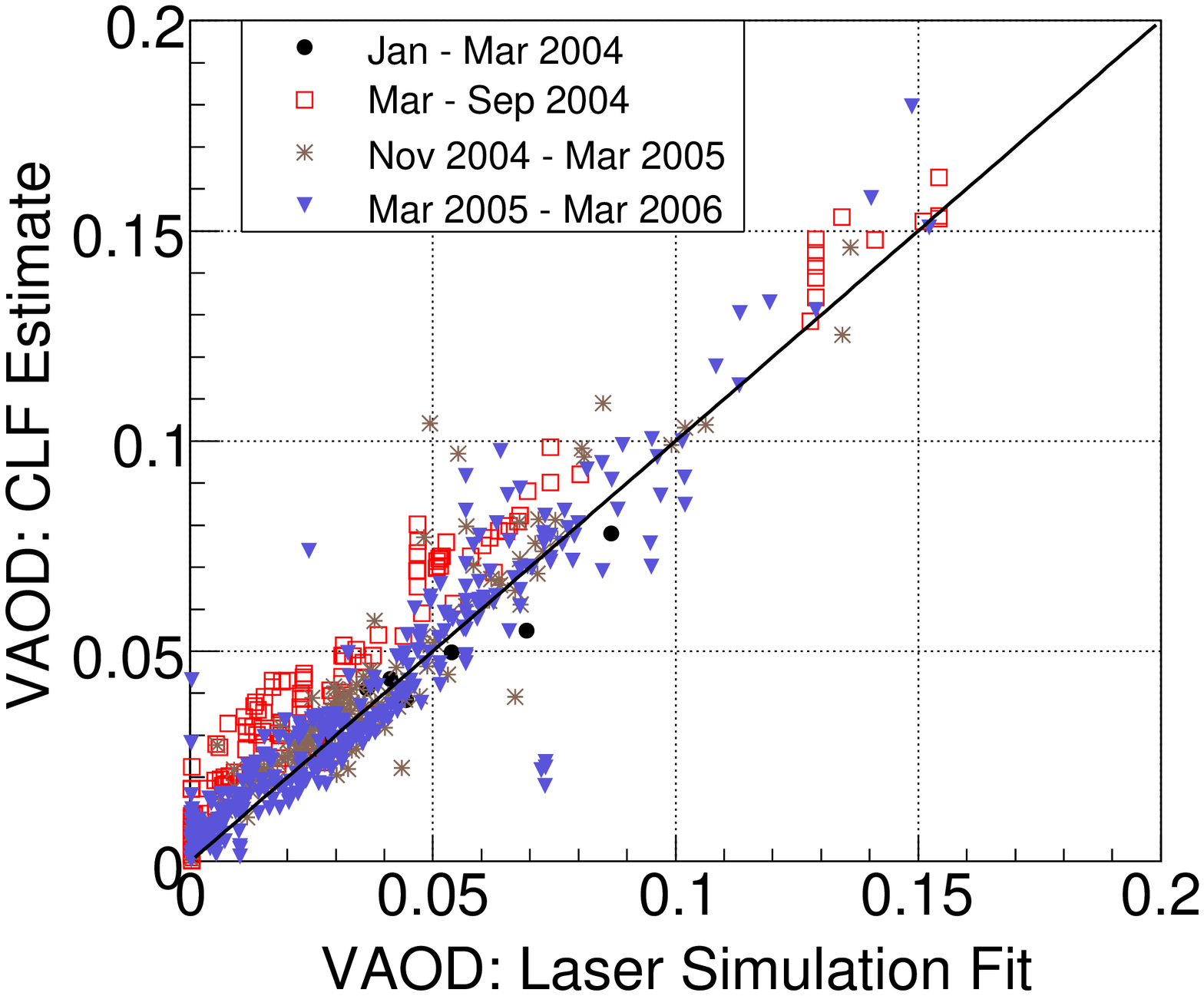}
        \caption{\label{fig:clfLaserCorr} VAOD at $4.5$~km from the CLF 
                 in different calibration epochs, compared to laser
                 simulations.}
        \end{center}
      \end{minipage}
      \hfill
      \begin{minipage}[t]{0.49\textwidth}
        \begin{center}
        \includegraphics*[width=0.75\textwidth,clip]{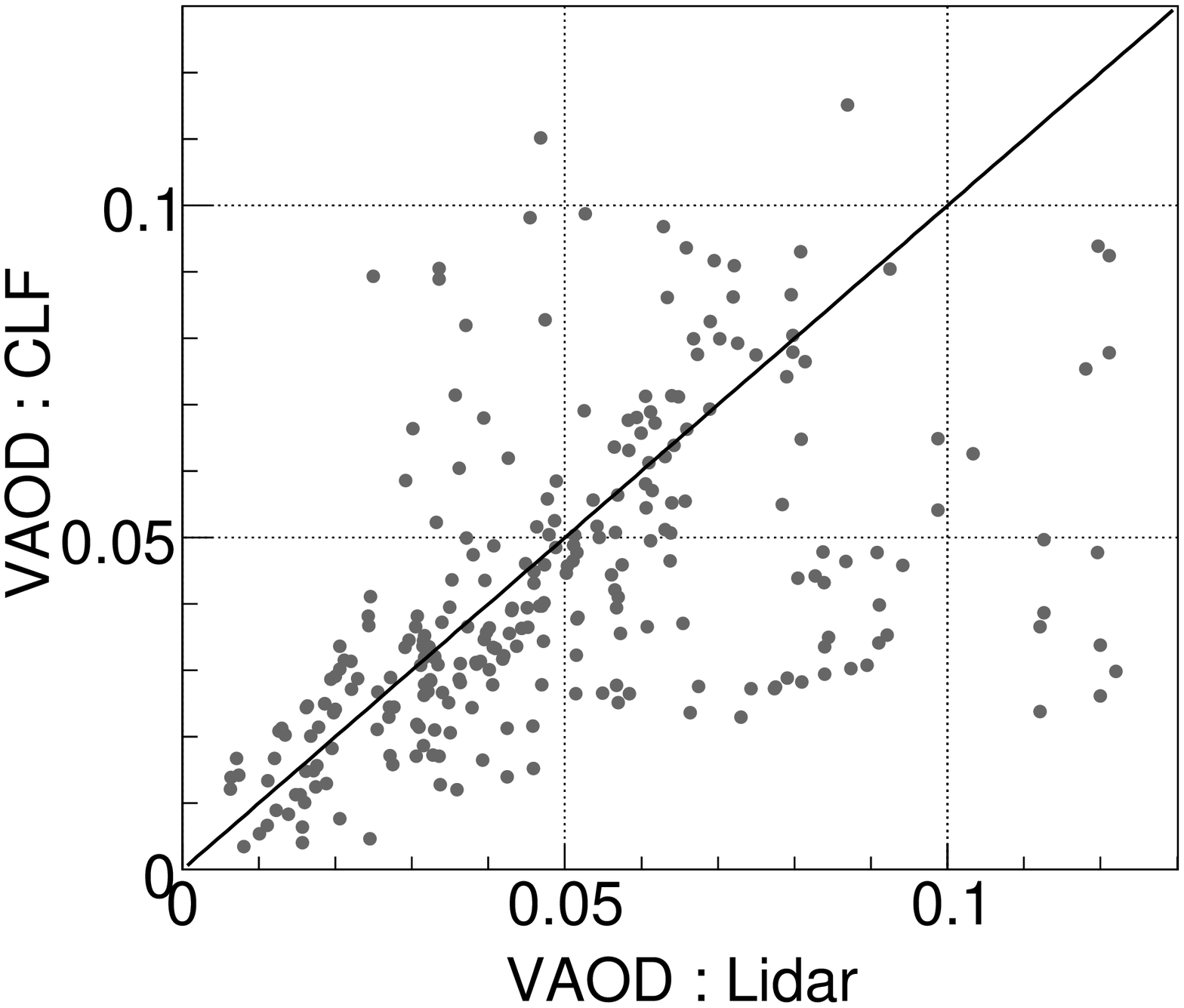}
        \caption{\label{fig:clfLidar} VAOD at $4.5$~km at Coihueco, Oct 2006
                 -- Jan 2007, observed by the CLF and Lidar.}
        \end{center}
      \end{minipage}
    \end{figure*}

    Every hour, each lidar sweeps the sky in a set pattern, pulsing the laser
    at 333~Hz and observing the backscattered light with its optical receivers.
    The sweeps occur outside the FD field of view to avoid triggering the
    detector.  By observing the backscattered light, the lidar can determine
    $\alpha(h)$ and $\tau(h)$ using a straightforward numerical
    inversion~\cite{Filipcic:2002ba}.  As shown in Figure~\ref{fig:clfLidar},
    the lidar and CLF results are in reasonable agreement, despite large
    differences in operation, analysis, and viewing regions.

    The lidar is also well suited to detect cloud layers, which create
    significant echoes in the backscattered light signal; these can be
    automatically detected by a simple gradient/threshold algorithm.  Under
    certain circumstances, the lidar can also estimate cloud optical
    depths~\cite{BenZvi:2006xb}.  The lidar is currently accumulating an
    extensive hourly database of cloud height, sky coverage, and optical depth.

\section{Scattering Measurements}

  The FD reconstruction must not only correct for the attenuation of
  scintillation light, but also subtract Cherenkov light scattered into the FD
  field of view.  Therefore, the scattering properties of the atmosphere should
  be well-understood.  Aerosol scattering is highly nontrivial and depends on
  the physical properties of the aerosols, but its distribution as a function
  of scattering angle $\theta$ can be approximated by the simple
  parameterization
  
  \begin{align}\label{eq:phase_function}
    P(\theta) &= \frac{1-g^2}{4\pi}\cdot\\
              & \left(\frac{1}{(1+g^2-2g\cos{\theta})^{3/2}}+
                f\frac{3\cos^2{\theta}-1}{2(1+g^2)^{3/2}}
                \right)\nonumber
  \end{align}

  where $g=\langle\cos{\theta}\rangle$ measures the asymmetry of
  scattering and $f$ determines the relative strength of forward and
  backward scattering.  The quantities $f$ and $g$ determine $P(\theta)$, and
  are affected by local aerosol characteristics.

  In the Auger Observatory, $f$ and $g$ are measured by Aerosol Phase Function
  monitors (APFs) located several km from the FDs at Coihueco and Los
  Morados \cite{BenZvi:2007px}.  Using a collimated xenon flash lamp, each APF
  fires an hourly sequence of $350$~nm shots horizontally across the FD field
  of view, covering $30^\circ$ to $150^\circ$ in azimuth.  The scattering
  parameters $f$ and $g$ can be determined simply by fitting the horizontal
  light track recorded by the FD.

  \begin{figure}[ht]
    \begin{center}
      \includegraphics*[width=0.47\textwidth,clip]{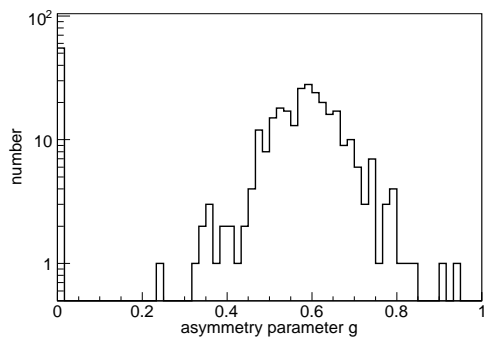}
      \caption{\label{fig:pf_params} Scattering parameter $g$
      measured at Coihueco between June 2006 and March 2007.}
    \end{center}
  \end{figure}

  Ten months of APF measurements at Coihueco have yielded a site average of
  $g=0.59\pm0.08$ for the local asymmetry parameter, excluding clear nights
  when $g=0$.  The distribution of $g$, shown in Figure~\ref{fig:pf_params}, is
  comparable to measurements reported in the literature for similar climates
  \cite{2006JGRD..11105S04A}.

\section{Wavelength Dependence}

  The attenuation of light by aerosols is expected to have some wavelength
  dependence, and this is typically parameterized by a power law

  \begin{equation}\label{eq:vaod_wl_dep}
    \tau(\lambda) = \tau_0\cdot
    \left(\frac{\lambda_0}{\lambda}\right)^\gamma
  \end{equation}

  In this expression, $\tau_0$ is the optical depth measured at the reference
  wavelength $\lambda_0=355$~nm, and $\gamma$ is the so-called Angstrom
  exponent of the dependence ($\gamma\approx4$ for molecular scattering).  

  \begin{figure}[ht]
    \begin{center}
      \includegraphics*[width=0.45\textwidth,clip]{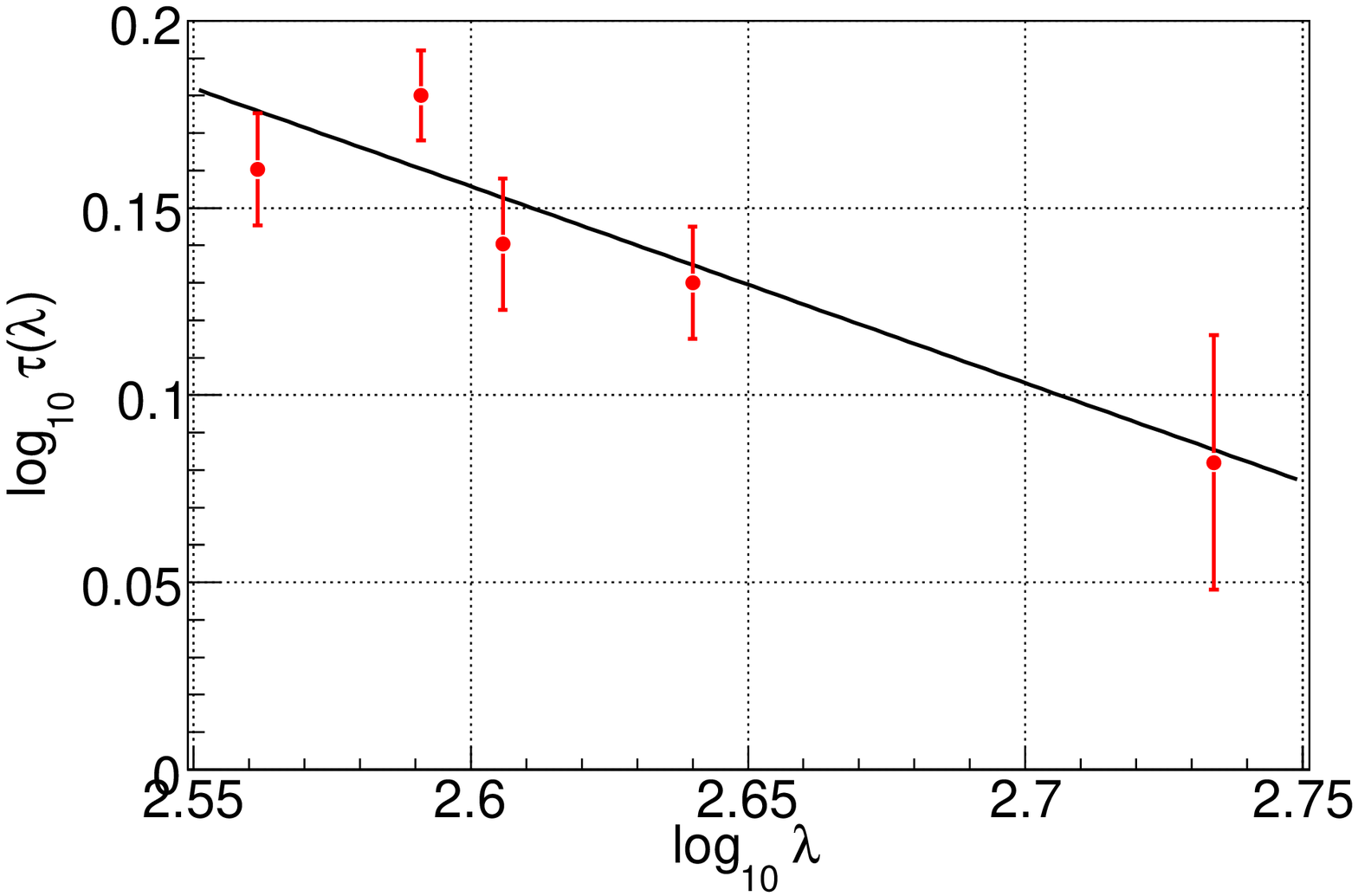}
      \caption{\label{fig:hamfit} HAM fit to CCD response after molecular
               subtraction for one measurement set.}
      \includegraphics*[width=0.45\textwidth,clip]{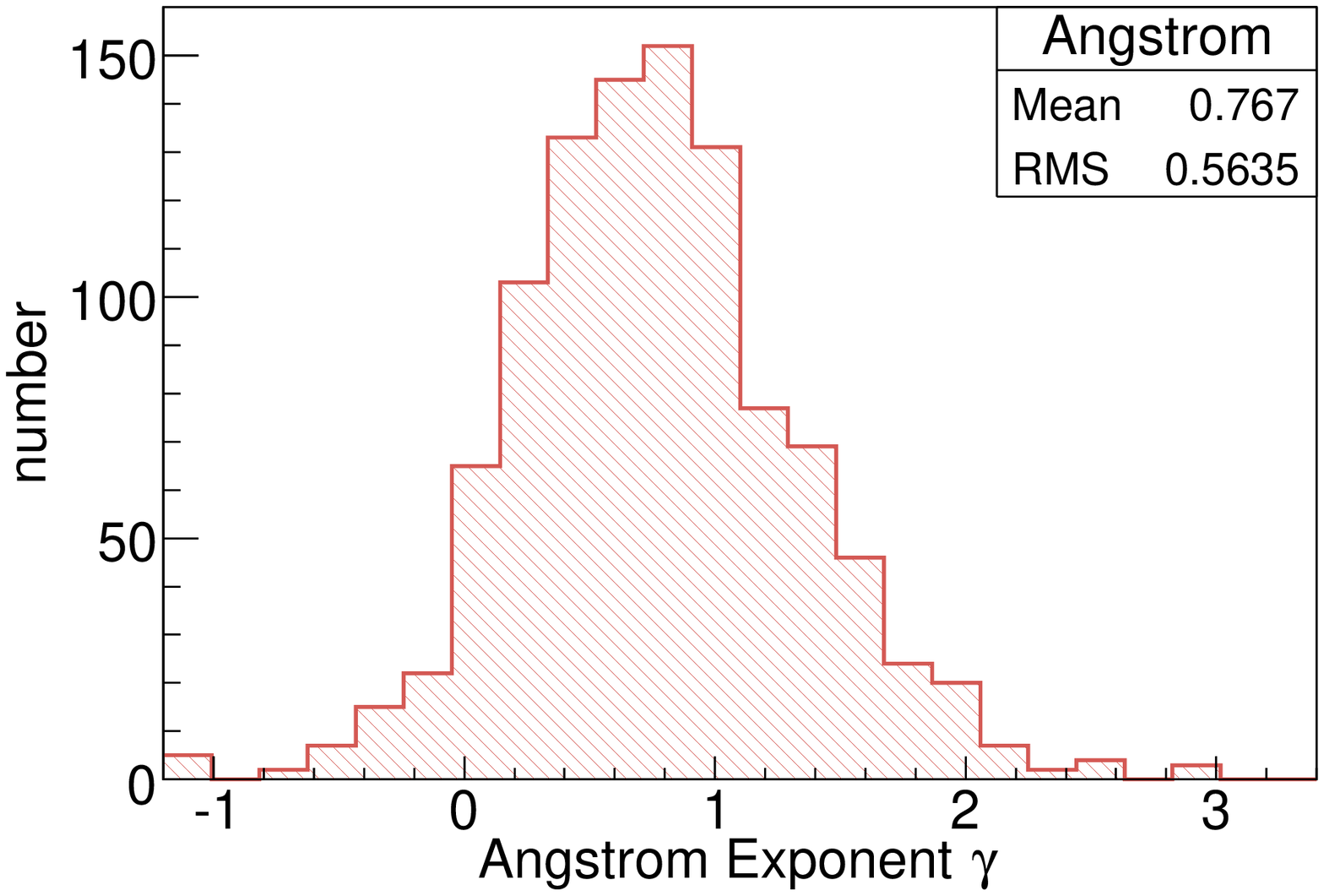}
      \caption{\label{fig:ham} Distribution of the Angstrom exponent observed
               by the HAM, July 2006 - February 2007.}
    \end{center}
  \end{figure}

  At the Auger Observatory, measurements of $\gamma$ are performed by two
  instruments: the robotic telescope FRAM, described in detail
  in~\cite{Travnicek:2007}; and the Horizontal Attenuation Monitor (HAM).  The
  HAM consists of a high intensity discharge lamp located at Coihueco, viewed
  by a CCD camera placed $\sim45$~km away at Los Leones.  Using a
  filter wheel, the camera records the aerosol extinction coefficient, and
  hence the aerosol optical depth between the two sites, at five different
  wavelengths. 

  Figure~\ref{fig:hamfit} shows a typical HAM fit used to estimate $\gamma$.
  The uncertainties are dominated by measurement fluctuations, and include a
  systematic effect due to subtraction of the estimated molecular optical depth
  between Los Leones and Coihueco.  The distribution of observed $\gamma$
  values, plotted in Fig.~\ref{fig:ham}, is consistent with other measurements
  and physical expectations~\cite{Travnicek:2007,Kaskaoutis:2006}.

\section{Impact of Aerosols on Shower Measurements}
  
  The measurement uncertainties in the VAOD, phase function, and wavelength
  dependence of aerosol attenuation have been propagated in the reconstruction
  of a select number of high-quality air showers~\cite{Prouza:2007}.  The
  dominant uncertainty comes from the VAOD, which contributes $5.5\%$ to the
  uncertainty in shower energy, while the phase function and wavelength
  dependence both contribute $\sim1\%$ to the uncertainty in the energy.  The
  effect of VAOD observations on $X_\mathrm{max}$, the depth of shower
  maximum, is $4\ \mathrm{g\ cm^{-2}}$, while the wavelength dependence and the
  phase function contribute uncertainties of $1$ and $2\ \mathrm{g\ cm^{-2}}$,
  respectively.

\bibliography{icrc0399}

\begin{thebibliography}{1}

\bibitem{2006JGRD..11105S04A}
E.~Andrews et~al.
\newblock {\em J. Geophys. Res.}, 111(D10):D05S04, 2006.

\bibitem{BenZvi:2006xb}
S.~Y. BenZvi et~al.
\newblock {\em Nucl. Instrum. Meth.}, A574:171--184, 2007.

\bibitem{BenZvi:2007px}
S.~Y. BenZvi et~al.
\newblock 2007.
\newblock arXiv:0704.0303 [astro-ph].

\bibitem{Fick:2006yn}
B.~Fick et~al.
\newblock {\em JINST}, 1:P11003, 2006.

\bibitem{Filipcic:2002ba}
A.~Filip\v{c}i\v{c} et~al.
\newblock {\em Astropart. Phys.}, 18:501--512, 2003.

\bibitem{Kaskaoutis:2006}
D.~G. Kaskaoutis et~al.
\newblock {\em J. Atmos. Solar-Terr. Phys.}, 68:2147, 2006.

\bibitem{Prouza:2007}
M.~Prouza.
\newblock In {\em Proc. 30th ICRC}, M\'{e}rida, M\'{e}xico, 2007.

\bibitem{Travnicek:2007}
P.~Tr\'{a}vn\'{\i}\v{c}ek.
\newblock In {\em Proc. 30th ICRC}, M\'{e}rida, M\'{e}xico, 2007.

\bibitem{Wiencke:2006hg}
Lawrence Wiencke.
\newblock {\em Nucl. Instrum. Meth.}, A572:508--510, 2007.

\end{thebibliography}
\bibliographystyle{plain}

\end{document}